\documentclass[twoside]{article}
\usepackage{amsmath,amsfonts,bm}

\usepackage[eqsecnum]{cmpj2e}

\title{On the spectral relations for multitime correlation functions}

\author{A.M. Shvaika}
\address{Institute for Condensed Matter Physics of the National Academy of Sciences of Ukraine,\\
1 Svientsitskii Str., 79011 Lviv, Ukraine}

\begin{document}

\maketitle

\begin{abstract}
A general approach for derivation of the spectral relations for the multitime correlation functions is presented. A special attention is paid to the consideration of the non-ergodic (conserving) contributions and it is shown that such contributions can be treated in a rigorous way using multitime temperature Green functions.
Representation of the multitime Green functions in terms of the spectral densities and solution of the reverse problem~--- finding of the spectral densities from the known Green functions are given for the case of the three-time correlation functions.
\pacs 05.30.-d 
\end{abstract}

\section{Introduction}

One of the main tasks of the quantum statistic physics is calculation of the correlation functions for many-body systems of different kind because they contain the most important information about the observable quantities and system properties. As it was first noticed by Kubo~\cite{kubo:570}, linear transport coefficients are expressed in terms of the Fourier transforms of appropriate correlation functions, which relate by spectral relations to the two-time Green functions. Since that time the Green's function method has been noticed and extensively developed~\cite{bogolyubov:53,zubarev:71,bonch-bruevich:book}, that provides equations which allow to calculate the expectation values of operators and observable quantities. But very soon it was noticed that spectral relations must be completed by the special treatment of the pole at zero frequency, which gives additional contribution connected with presence of the conserving quantities \cite{stevens:1307,suzuki:882,suzuki:277}. Later, it was shown that such contributions describe the difference between the isothermal and isolated response of the many-body system and are specific for the non-ergodic systems~\cite{kubo:570,suzuki:882,suzuki:277,wilcox:624} where the regions in the phase space which can not be achieved by the trajectory of the point that describes an evolution of many-body system exist. In the Green's function formalism the issue of ergodicity appears as a difficulty in the determination of the zero-frequency bosonic propagators~\cite{stevens:1307,suzuki:882,suzuki:277,fernandez:505,callen:505,lucas:503,morita:1030,kwok:1196,ramos:441,huber:199,aksenov:K43,aksenov:56,aksenov:375}.
Nevertheless, even now many textbooks on the quantum statistics and many-body theory do not provide complete discussion of the spectral relations and special treatment of the zero-frequency functions.

The equations of motion do not uniquely determine the causal and retarded Green's functions, but only up to $\delta$-function of frequency with some unknown coefficients which produce additional contributions in the zero-frequency functions (see, section~\ref{2time}). Usually, these zero-frequency functions are fixed by assigning them their ergodic values, but this cannot be justified a priori. A wrong determination of them dramatically affects the values of directly measurable quantities like compressibility, specific heat, and magnetic susceptibility. In order to handle this zero-frequency functions a different approaches were developed, e.g. anti-commutator bosonic Green functions~\cite{ramos:441}, direct algebraic method~\cite{sarry:47}, singular-value decomposition~\cite{frobrich:014410}, algebra constraints in the composite operator method~\cite{mancini:37,mancini:537}, etc. 

On the other hand, temperature Green functions~\cite{matsubara:351,abrikosov:book} are free from this issues and allow to avoid all complications connected with the presense of the non-ergodic terms~\cite{kwok:1196}, e.g. they allow to calculate isothermal susceptibilities for the Ising model, where Kubo response is equal to zero, as well as for the more complicated spin and electron~\cite{baryakhtar:book,izyumov:book} and pseudospin-electron~\cite{stasyuk:134} systems, and for the infinite-dimensional Falicov-Kimball model~\cite{falicov:997,freericks:1333}, where exact expression for the isothermal charge susceptibility~\cite{shvaika:177,shvaika:349} contains both Kubo response, which is finite at all temperatures, and non-ergodic contribution, whose divergencies give the phase transition points.

The mentioned above issues about special treatment of the zero-frequency functions are mostly investigated for the two-time correlation and Green functions, but nobody have cosidered this for the multitime one. In the original Kubo's paper~\cite{kubo:570}, the formulation of the transport theory is not limited to the linear phenomena and solution of the Liouville equation for the density matrix is given to the arbitrary order in the strength of disturbance. Resulting multitime correlation and Green functions can be used for the description of the nonlinear transport phenomena and resonances~\cite{tyablikov:142,tanaka:388,harper:1613}. Besides, multitime correlation functions also appear as puzzles in different orders of the perturbation theories for many-body systems~\cite{shvets:438}. Moreover, cross-sections of the inelastic scattering processes can be expressed in terms of the multitime correlation functions too, e.g. for the electronic inelastic light (Raman) scattering the nonresonant, mixed, and resonant responses are connected with the two-time, three-time, and four-time temperature Green functions~\cite{shvaika:137402,shvaika:045120}, respectively, and can be rewritten in terms of the multitime correlation functions. Nonresonant contribution is connected with the spectral density of the two-time correlation function
\begin{align}\label{Raman_N_2t}
    R_N(\bm q,\Omega) = 2\pi g^2(\bm k_i) g^2(\bm k_o) I_{\tilde \gamma\tilde \gamma}(\Omega,-\Omega),
\end{align}
mixed contribution is connected with the spectral density of the three-time correlation functions
\begin{align}\label{Raman_M_3t}
    R_M(\bm q,\Omega) &{}= 2\pi g^2(\bm k_i) g^2(\bm k_o)
     \int\limits_{-\infty}^{+\infty}\rd\omega
    \left[
    \frac{I_{\Tilde\gamma j^{(o)} j^{(i)}}(\Omega,\omega-\Omega,-\omega)}
    {\omega - \omega_i + \ri\delta}
    \right.
    \\
    \nonumber
        &{} +
    \left.
    \frac{I_{\Tilde \gamma j^{(i)} j^{(o)}} (\Omega,\omega-\Omega,-\omega)}
    {\omega + \omega_o - \ri\delta}
    +
    \frac{I_{j^{(i)} j^{(o)} \Tilde \gamma}(\omega,\Omega-\omega,-\Omega)}
    {\omega - \omega_i - \ri\delta}
    +
    \frac{I_{j^{(o)} j^{(i)} \Tilde \gamma}(\omega,\Omega-\omega,-\Omega)}
    {\omega + \omega_o + \ri\delta}
    \right],
\end{align}
and resonant contribution is connected with the spectral density of the four-time one
\begin{align}\label{Raman_R_4t}
    R_R(\bm q,\Omega) &= 2\pi g^2(\bm k_i) g^2(\bm k_o)
     \int\limits_{-\infty}^{+\infty}\rd\omega
     \int\limits_{-\infty}^{+\infty}\rd\Tilde\omega
    \\
    \nonumber
    &\times\left[
    \frac{I_{j^{(i)} j^{(o)} j^{(o)} j^{(i)}} (\omega,\Omega-\omega,\tilde\omega-\Omega,-\tilde\omega)}
    {
    (\omega - \omega_i - \ri\delta)
    (\tilde\omega - \omega_i + \ri\delta)
    } +
    \frac{I_{j^{(o)} j^{(i)} j^{(i)} j^{(o)}} (\omega,\Omega-\omega,\tilde\omega-\Omega,-\tilde\omega)}
    {
    (\omega + \omega_o + \ri\delta)
    (\tilde\omega + \omega_o - \ri\delta)
    }
    \right.
     \\
     \nonumber
     &{}+
    \left.
    \frac{I_{j^{(o)} j^{(i)} j^{(o)} j^{(i)}} (\omega,\Omega-\omega,\tilde\omega-\Omega,-\tilde\omega)}
    {
    (\omega + \omega_o + \ri\delta)
    (\tilde\omega - \omega_i + \ri\delta)
    } +
    \frac{I_{j^{(i)} j^{(o)} j^{(i)} j^{(o)}} (\omega,\Omega-\omega,\tilde\omega-\Omega,-\tilde\omega)}
    {
    (\omega - \omega_i - \ri\delta)
    (\tilde\omega + \omega_o - \ri\delta)
    }
    \right] ,
\end{align}
where $I_{AB\ldots}(\omega_1,\omega_2,\ldots)$ are spectral densities of multitime correlation functions (see below) and $\Hat{\Tilde\gamma}$ and $\hat j^{(i,o)}$ are expressed in terms of the stress tensor and current operator, respectively.

In general, spectral relations include (a) representation of the observable quantities in terms of the spectral densities of the multitime correlation functions [like Eqs.~(\ref{Raman_N_2t}), (\ref{Raman_M_3t}), and (\ref{Raman_R_4t}) for the inelastic light scattering], (b) extraction of the non-ergodic contributions in the spectral densities, (c) representation of the multitime temperature Green functions in terms of the spectral densities, and (d) solution of the reverse problem~--- finding of the spectral densities from the known multitime temperature Green functions and calculation on this basis of the observable quantities.
Spectral relations for multitime correlation functions were formulated in Ref.~\cite{bonch-bruevich:book,bonch-bruevich:529} only for the case of the multitime generalizations of the retarded and advanced functions without any consideration of the zero-frequency non-ergodic contributions. The main purpose of this paper is to show how non-ergodic contributions enter in multitime correlation functions and how a complete set of the spectral relations for the multitime correlation and Green functions can be derived.

\section{Two-time correlation functions}\label{2time}

Before we start consideration of the spectral relations for the multitime correlation functions, let us remind basic relations and main results for the two-time correlation functions. In general, two-time correlation function is defined by the following expression
\begin{equation}\label{CF2}
  K_{AB}(t_1-t_2)=\langle \hat A(t_1) \hat B(t_2)\rangle,
\end{equation}
where $\hat A(t)=\re^{\ri Ht}\hat A\re^{-\ri Ht}$ is an operator $\hat A$ in the Heisenberg representation, and angular brackets denote statistical average with total Hamiltonian of the many-body system. Here and below we shall consider the case of the bosonic operators only, generalization for the fermionic one is obvious. We can perform Fourier transformation of Eq.~(\ref{CF2}) that gives a spectral density for the two-time correlation function
\begin{align}\label{CF2FT}
  I_{AB}(\omega_1,-\omega_1)&{}= \frac{1}{2\pi} \!\int\limits_{-\infty}^{+\infty}\!\!
                  \rd (t_1-t_2) \re^{i\omega_1(t_1-t_2)}K_{AB}(t_1-t_2)
       \\ \nonumber
            &{}= \tilde I_{AB}(\omega_1,-\omega_1) + \delta(\omega_1) \tilde I_{AB}(\circ,\circ),
\end{align}
where we have separated two contributions
\begin{align}\label{CF2FT1}
  \tilde I_{AB}(\omega_1,-\omega_1)&{}=\frac{1}{\mathcal Z}
         {\sum_{jl}}'\re^{-\beta\varepsilon_j} A_{jl}B_{lj}\delta(\varepsilon_{jl}+\omega_1),\\
  \tilde I_{AB}(\circ,\circ)&{}=\frac{1}{\mathcal Z}
         \sum_{\substack{jl\\ \varepsilon_j=\varepsilon_l}}\re^{-\beta\varepsilon_j} A_{jl}B_{lj}
   \label{CF2FT2}
\end{align}
with different frequency and time dependences. The first one is time-dependent and includes sum over states with different energies ($\varepsilon_{jl}\equiv\varepsilon_j-\varepsilon_l\ne0$) denoted by prime (\ref{CF2FT1}), we shall call it regular contribution, and the second one is purely static and includes sum over the states with the same energy (\ref{CF2FT2}), we shall call it non-ergodic one. Here, $A_{jl}=\langle j| \hat A|l\rangle$ are matrix elements of the operator $\hat A$, $\mathcal{Z}=\sum_l\re^{-\beta\varepsilon_l}$ is partition function, and $\circ$ denotes that given contribution does not depend on the respective frequency. Besides, spectral densities (\ref{CF2FT}) satisfy the following permutation relation
\begin{equation}
  I_{AB}(\omega_1,-\omega_1)=I_{BA}(-\omega_1,\omega_1)\re^{\beta\omega_1}
\end{equation}
and for the given operators $\hat A$ and $\hat B$ there is only one nonidentical spectral density. At this point there are no special need to separate regular and non-ergodic contributions, and, as a rule, in the textbooks nobody do this. But such separation becomes important when one is going to find spectral densities using spectral relations for the two-time retarded Green's function
\begin{equation}
   G_{AB}^{(r)}(t_1-t_2)=\langle\langle\hat A(t_1)|\hat B(t_2)\rangle\rangle
                 =-\ri\Theta(t_1-t_2)\langle[\hat A(t_1),\hat B(t_2)]\rangle.
\end{equation}
Its Fourier transformation, from the formal point of view,
\begin{align}
  G_{AB}(\omega_1,-\omega_1)&{}= \dfrac1{2\pi}\!\int\limits_{-\infty}^{\infty}\!\!
                  \rd(t_1-t_2) \re^{\ri\omega_1(t_1-t_2)}G_{AB}(t_1-t_2)
      \\ \nonumber
       &{}= \widetilde G_{AB}(\omega_1,-\omega_1) + \delta(\omega_1) \widetilde G_{c}(\circ,\circ)
\end{align}
also includes two contributions with different time dependences
\begin{align}
  \widetilde G_{AB}(\omega_1,-\omega_1) &{}= \int\limits_{-\infty}^{+\infty} \rd\tilde\omega_1
            \tilde I_{AB}(\tilde\omega_1,-\tilde\omega_1)
            \frac{1-\re^{-\beta\tilde\omega_1}}{\tilde\omega_1-\omega_1\pm\ri\delta},  \\
  \widetilde G_{AB}(\circ,\circ) &{}= \tilde I_{AB}(\circ,\circ),
\end{align}
but only first one can be derived using an equation of motion technics
\begin{equation}
    \omega_1 G_{AB}(\omega_1,-\omega_1)= \omega_1\widetilde G_{AB}(\omega_1,-\omega_1)
\end{equation}
and the second one is omitted in the equations of motion
\begin{equation}
                  \omega_1\delta(\omega_1)\widetilde G_{AB}(\circ,\circ)\equiv0.
    \nonumber
\end{equation}
As a result, a corresponding static contribution into spectral density can not be handled directly by the spectral theorem for the retarded Green functions.

On the other hand, a method of the temperature or Matsubara Green functions allows to avoid all complications connected with the presense of the non-ergodic terms and obtain all contributions in the spectral density in a straightforward way~\cite{kwok:1196}. In general, a two-time temperature Green's function can be defined as
\begin{equation}
  K_{c}(\tau_1-\tau_2)=\langle \mathcal T \hat A(\tau_1) \hat B(\tau_2)\rangle,
\end{equation}
where $\hat A(\tau)=\re^{H\tau}\hat A\re^{-H\tau}$, $\tau$ is imaginary time (inverse temperature), and $\mathcal T$ is operator of the imaginary time chronological ordering. Its Fourier transform
\begin{align}
  K_{c}(\ri\nu_1,-\ri\nu_1)&{}= \int\limits_{0}^{\beta}
                  \rd(\tau_1-\tau_2) \re^{\ri\nu_1(\tau_1-\tau_2)}K_{c}(\tau_1-\tau_2)
     \\ \nonumber
      &{}= \widetilde K_{c}(\ri\nu_1,-\ri\nu_1) + \beta\Delta(\ri\nu_1) \widetilde K_{c}(\circ,\circ),
\end{align}
where $\ri\nu_1\equiv \ri\omega_{\nu_1}=2\pi\ri T \nu_1$ are Matsubara's frequency and
\begin{equation}
   \Delta(z)=\left\{\begin{array}{ll}
   1, \quad z=0\\
   0, \quad z\neq 0
   \end{array}
   \right.
\end{equation}
is generalization of the Kronecker symbol, also contains two contributions with different time dependences
\begin{align}
  \widetilde K_{c}(\ri\nu_1,-\ri\nu_1) &{}= \int\limits_{-\infty}^{+\infty} \rd\tilde\omega_1
            \tilde I_{AB}(\tilde\omega_1,-\tilde\omega_1)
            \frac{1-\re^{-\beta\tilde\omega_1}}{\tilde\omega_1-\ri\nu_1},  \\
   \label{2TGFcontrb}
  \widetilde K_{c}(\circ,\circ) &{}= \tilde I_{AB}(\circ,\circ).
\end{align}
As a rule, temperature Green functions are handled using different kinds of the diagrammatic technics which allow to calculate contributions with $\Delta$-symbols and according to (\ref{2TGFcontrb}) they can be identified as a non-ergodic contributions in spectral densities, that is the main difference with the case of the retarded Green functions. For the non-ergodic systems such contributions are the main one which determine critical behavior. A typical example is the Ising model for which there are only non-ergodic contributions whose divergencies give Curie points and there are no regular one. Another example of the non-ergodic fermionic system is the Falicov-Kimball model~\cite{falicov:997,freericks:1333} whose isothermal charge susceptibility contains both non-ergodic and regular contributions but only the first one determines the critical point~\cite{shvaika:177,shvaika:349}. Regular contribution in spectral density is connected with nonanalyticy (imaginary part) of the Green's function at real axis and can be obtained by performing an analytic continuation of the Matsubara Green's function from the imaginary to complex frequencies and then to real one $\ri\nu_1 \to z_1 \to \omega_1\pm \ri\delta$
\begin{equation}
  \widetilde K_{c}(z_1,-z_1)\biggr|_1 \equiv \frac{1}{2\pi \ri}
  \widetilde K_{c}(z_1,-z_1)\biggr|_{z_1=\omega_1-\ri\delta}^{z_1=\omega_1+\ri\delta}
  = \tilde I_{AB}(\omega_1,-\omega_1) (1-\re^{-\beta\omega_1}),
\end{equation}
that completes a spectral relations for the two-time correlation functions.

\section{Three-time correlation functions}

Now let us proceed to the consideration of the three-time correlation functions. Generalization for the case of the higher-order multitime correlation functions can be done in the same way, but it is much more cumbersome and will not be considered here. 

Three-time correlation function can be defined in a usual way as
\begin{equation}
  K_{ABC}(t_1,t_2,t_3)=\langle \hat A(t_1) \hat B(t_2) \hat C(t_3)\rangle.
\end{equation}
Here, we shall consider only the case of the equilibrium many-body systems which are time-shift invariant
\begin{equation}
  K_{ABC}(t_1,t_2,t_3)=K_{ABC}(t_1-t,t_2-t,t_3-t).
\end{equation}

Spectral density is defined as its Fourier transform
\begin{align}\label{SD_3t}
  I_{ABC}(\omega_1,\omega_2,\omega_3) &{}= \frac{1}{(2\pi)^2}
        \int\limits_{-\infty}^{+\infty}\rd (t_1-t_3)
        \int\limits_{-\infty}^{+\infty}\rd (t_2-t_3)
            \re^{i(\omega_1 t_1 + \omega_2 t_2 + \omega_3 t_3)}K_{ABC}(t_1,t_2,t_3)\\
            &{}= \bigl[\tilde I_{ABC}(\omega_1,\omega_2,\omega_3) 
             + \delta(\omega_1) \tilde I_{ABC}(\circ,-\omega_3,\omega_3)
            + \delta(\omega_2) \tilde I_{ABC}(\omega_1,\circ,-\omega_1)
            \nonumber\\
            &{}
            + \delta(\omega_3) \tilde I_{ABC}(-\omega_2,\omega_2,\circ)
            + \delta(\omega_1)\delta(\omega_2) \tilde I_{ABC}(\circ,\circ,\circ) 
                \bigr]
            \Delta(\omega_1+\omega_2+\omega_3)
            \nonumber
\end{align}
and includes five different contributions with different time dependences
\begin{align}
  \tilde I_{ABC}(\omega_1,\omega_2,-\omega_1-\omega_2)&{}=\frac{1}{\mathcal Z}
         {\sum_{jlf}}'\re^{-\beta\varepsilon_j} A_{jl}B_{lf}C_{fj}
                       \delta(\varepsilon_{jl}+\omega_1)\delta(\varepsilon_{lf}+\omega_2),
        \label{SD_3t_cont}\\
  \tilde I_{ABC}(\circ,-\omega_3,\omega_3)&{}=\frac{1}{\mathcal Z}
         \sum_{\substack{jlf\\ \varepsilon_j=\varepsilon_l\neq\varepsilon_f}}\re^{-\beta\varepsilon_j} 
                    A_{jl}B_{lf}C_{fj}\delta(\varepsilon_{fj}+\omega_3),
        \label{SD_3t_cont1}\\
  \tilde I_{ABC}(\omega_1,\circ,-\omega_1)&{}=\frac{1}{\mathcal Z}
         \sum_{\substack{jlf\\ \varepsilon_l=\varepsilon_f\neq\varepsilon_j}}\re^{-\beta\varepsilon_j} 
                    A_{jl}B_{lf}C_{fj}\delta(\varepsilon_{jl}+\omega_1),
        \label{SD_3t_cont2}\\
  \tilde I_{ABC}(-\omega_2,\omega_2,\circ)&{}=\frac{1}{\mathcal Z}
         \sum_{\substack{jlf\\ \varepsilon_f=\varepsilon_j\neq\varepsilon_l}}\re^{-\beta\varepsilon_j} 
                    A_{jl}B_{lf}C_{fj}\delta(\varepsilon_{lf}+\omega_2),
        \label{SD_3t_cont3}\\
  \tilde I_{ABC}(\circ,\circ,\circ)&{}=\frac{1}{\mathcal Z}
         \sum_{\substack{jlf\\ \varepsilon_f=\varepsilon_j=\varepsilon_l}}\re^{-\beta\varepsilon_j} 
                    A_{jl}B_{lf}C_{fj}.
        \label{SD_3t_cont0}
\end{align}
Besides, the total spectral density (\ref{SD_3t}) as well as each contribution satisfy the following cyclic permutation identities ($\omega_1+\omega_2+\omega_3=0$)~\cite{bonch-bruevich:book,shvets:438}
\begin{equation}\label{CF3tCicle}
    I_{ABC}(\omega_1,\omega_2,\omega_3)
    =I_{BCA}(\omega_2,\omega_3,\omega_1)\re^{\beta\omega_1}
    =I_{CAB}(\omega_3,\omega_1,\omega_2)\re^{-\beta\omega_3}
\end{equation}
and for the given operators $\hat A$, $\hat B$, and $\hat C$ there are only two nonidentical spectral densities, e.g. $I_{ABC}(\omega_1,\omega_2,\omega_3)$ and $I_{CBA}(\omega_3,\omega_2,\omega_1)$.

Now we introduce three-time temperature Green's function
\begin{align}
  K_{c}(\tau_1,\tau_2,\tau_3)&=\langle \mathcal T \hat A(\tau_1) \hat B(\tau_2) \hat C(\tau_3)\rangle,\\
  K_{c}(\tau_1,\tau_2,\tau_3)&=K_{c}(\tau_1-\tau,\tau_2-\tau,\tau_3-\tau).
  \nonumber
\end{align}
Due to the imaginary time ordering its Fourier transform contains $3!=6$ terms which can be collected into two groups of three terms connected by the cyclic permutations
\begin{align}\label{K3tFT}
  K_{c}(\ri\nu_1,\ri\nu_2,\ri\nu_3)& {}= \frac{1}{\beta}\int\limits_{0}^{\beta}\rd\tau_1
   \int\limits_{0}^{\beta}\rd\tau_2 \int\limits_{0}^{\beta}\rd\tau_3
          \re^{(\ri\nu_1\tau_1+\ri\nu_2\tau_2+\ri\nu_3\tau_3)}K_{c}(\tau_1,\tau_2,\tau_3)\\
            &{} = \frac{1}{\mathcal{Z}}\sum_{jlf}\bigl[
            A_{jl}B_{lf}C_{fj}\mathfrak P(j,\ri\nu_1,l,\ri\nu_2,f,\ri\nu_3)
            + C_{jf}B_{fl}A_{lj}\mathfrak P(j,\ri\nu_3,f,\ri\nu_2,l,\ri\nu_1)
            \bigr],
            \nonumber
\end{align}
where
\begin{align}\label{P_int}
     &\mathfrak P(j,\ri\nu_1,l,\ri\nu_2,f,\ri\nu_3)\\
     &{}=\frac{1}{\beta}\biggl[
     \re^{-\beta\varepsilon_j}\int\limits_{0}^{\beta}\rd\tau_1 \int\limits_{0}^{\tau_1}\rd\tau_2
                               \int\limits_{0}^{\tau_2}\rd\tau_3
     +\re^{-\beta\varepsilon_l}\int\limits_{0}^{\beta}\rd\tau_2 \int\limits_{0}^{\tau_2}\rd\tau_3
                               \int\limits_{0}^{\tau_3}\rd\tau_1
     +\re^{-\beta\varepsilon_f}\int\limits_{0}^{\beta}\rd\tau_3 \int\limits_{0}^{\tau_3}\rd\tau_1
                               \int\limits_{0}^{\tau_3}\rd\tau_2
     \biggr]
     \nonumber\\
     &{}\times\exp[(\varepsilon_{jl}+\ri\nu_1)\tau_1+(\varepsilon_{lf}+\ri\nu_2)\tau_2
                   +(\varepsilon_{fj}+\ri\nu_3)\tau_3].
     \nonumber
\end{align}
In the general case, when all Matsubara frequencies are nonzero or when there are no eigenstates with the same energy value, function (\ref{P_int}) is equal
\begin{align}\label{P_int_gen}
     &\widetilde{\mathfrak P}(j,\ri\nu_1,l,\ri\nu_2,f,\ri\nu_3)\\
     &{}=\Delta(\ri\nu_1+\ri\nu_2+\ri\nu_3)\biggl[
     \dfrac{\re^{-\beta\varepsilon_j}}{(\varepsilon_{lj}-\ri\nu_1)(\varepsilon_{fj}+\ri\nu_3)}
     + \dfrac{\re^{-\beta\varepsilon_l}}{(\varepsilon_{fl}-\ri\nu_2)(\varepsilon_{jl}+\ri\nu_1)}
     + \dfrac{\re^{-\beta\varepsilon_f}}{(\varepsilon_{if}-\ri\nu_3)(\varepsilon_{lf}+\ri\nu_2)}
     \biggr].
     \nonumber
\end{align}
Besides, we must consider several special cases, when we have levels with the same energy value: case $\varepsilon_{j}=\varepsilon_{l}\neq\varepsilon_{f}$ and $\ri\nu_1=-\ri\nu_2-\ri\nu_3=0$, when
\begin{equation}\label{P_int_1}
     \mathfrak P(j,0,l,\ri\nu_2,f,-\ri\nu_2)=\Delta(\ri\nu_2+\ri\nu_3)\biggl[
     \dfrac{\beta\re^{-\beta\varepsilon_l}}{\varepsilon_{fl}-\ri\nu_2}
     + \dfrac{\re^{-\beta\varepsilon_f}-\re^{-\beta\varepsilon_l}}{(\varepsilon_{fl}-\ri\nu_2)^2}
     \biggr],
\end{equation}
case $\varepsilon_{l}=\varepsilon_{f}\neq\varepsilon_{j}$ and $\ri\nu_2=-\ri\nu_3-\ri\nu_1=0$, when
\begin{equation}\label{P_int_2}
     \mathfrak P(j,-\ri\nu_3,l,0,f,\ri\nu_3)=\Delta(\ri\nu_3+\ri\nu_1)\biggl[
      \dfrac{\beta\re^{-\beta\varepsilon_f}}{\varepsilon_{jf}-\ri\nu_3}
      + \dfrac{\re^{-\beta\varepsilon_j}-\re^{-\beta\varepsilon_f}}{(\varepsilon_{jf}-\ri\nu_3)^2}
     \biggr],
\end{equation}
case $\varepsilon_{f}=\varepsilon_{j}\neq\varepsilon_{l}$ and $\ri\nu_3=-\ri\nu_1-\ri\nu_2=0$, when
\begin{equation}\label{P_int_3}
     \mathfrak P(j,\ri\nu_1,l,-\ri\nu_1,f,0)=\Delta(\ri\nu_1+\ri\nu_2)\biggl[
      \dfrac{\beta\re^{-\beta\varepsilon_j}}{\varepsilon_{lj}-\ri\nu_1}
      + \dfrac{\re^{-\beta\varepsilon_l}-\re^{-\beta\varepsilon_j}}{(\varepsilon_{lj}-\ri\nu_1)^2}
     \biggr],
\end{equation}
and case $\varepsilon_{j}=\varepsilon_{l}=\varepsilon_{f}$ and $\ri\nu_1=\ri\nu_2=\ri\nu_3=0$, when
\begin{equation}\label{P_int_0}
     \mathfrak P(j,0,l,0,f,0)=
      \dfrac{\beta^2}{2}\re^{-\beta\varepsilon_j}.
\end{equation}
The second term in the r.h.s. of Eq.~(\ref{P_int_1}) can be derived from Eq.~(\ref{P_int_gen}) by the analytic continuation of the Matsubara frequencies to the complex one $\ri\nu_1\to z_1$ and $\ri\nu_3\to -\ri\nu_2-z_1$ followed by the limit $z_1\to0$, but the first one can not be derived from Eq.~(\ref{P_int_gen}) by manipulating by frequencies only and corresponds to the additional non-ergodic or conserving contribution, which originates from the presence of the states with the same energies, and appears only at zero frequency. Such non-ergodic contributions can be derived from Eq.~(\ref{P_int_gen}) by
\begin{equation}
   \left[\lim_{\varepsilon_{jl}\to0}\lim_{z_1\to0}-\lim_{z_1\to0}\lim_{\varepsilon_{jl}\to0}\right]
\widetilde{\mathfrak P}(j,z_1,l,z_2,f,-z_1-z_2)=\dfrac{\beta\re^{-\beta\varepsilon_l}}{\varepsilon_{fl}-z_2},
\end{equation}
but such derivation involves manipulations with the many-body quantum states energies, that can not be, in general, reproduced by the quantum statistics many-body methods.
A similar analysis can be done for Eqs.~(\ref{P_int_2}) and (\ref{P_int_3}) and, after analytic continuation from the imaginary axis to the complex plane $\ri\nu_i\to z_i$ with using constraint
\begin{equation}\label{z_constr}
   \sum_i z_i=0,
\end{equation}
Eq.~(\ref{P_int}) can be rewritten as
\begin{align}\label{P_ac}
  \mathfrak P(j,z_1,l,z_2,f,z_3)&{}=\Delta(z_1+z_2+z_3)
       \biggl[\dfrac{\beta^2}{2}\Delta(z_1)\Delta(z_2)
               \Delta_{\varepsilon_j,\varepsilon_l,\varepsilon_f}\re^{-\beta\varepsilon_j}\\
     &{}+ \beta\Delta(z_1)\Delta_{\varepsilon_j,\varepsilon_l}
                \dfrac{\re^{-\beta\varepsilon_l}}{\varepsilon_{fl}-z_2}
     + \beta\Delta(z_2)\Delta_{\varepsilon_l,\varepsilon_f}
                \dfrac{\re^{-\beta\varepsilon_f}}{\varepsilon_{jf}-z_3}
     + \beta\Delta(z_3)\Delta_{\varepsilon_f,\varepsilon_j}
                \dfrac{\re^{-\beta\varepsilon_j}}{\varepsilon_{lj}-z_1}
     \nonumber\\
     &{}+ \dfrac{\re^{-\beta\varepsilon_j}}{(\varepsilon_{lj}-z_1)(\varepsilon_{fj}+z_3)}
     + \dfrac{\re^{-\beta\varepsilon_l}}{(\varepsilon_{fl}-z_2)(\varepsilon_{jl}+z_1)}
     + \dfrac{\re^{-\beta\varepsilon_f}}{(\varepsilon_{if}-z_3)(\varepsilon_{lf}+z_2)}
     \biggr],
     \nonumber
\end{align}
where
 \begin{equation}
     \Delta_{\varepsilon_j,\ldots,\varepsilon_f}=\left\{\begin{array}{ll}
    1, \quad \varepsilon_j=\ldots=\varepsilon_f\\
    0, \quad \text{other case}
    \end{array}
    \right..
 \end{equation}

Now, after substitution of Eq.~(\ref{P_ac}) in Eq.~(\ref{K3tFT}), we get the following representation for the analytically continued three-time temperature Green's function
\begin{align}\label{KcvsSD3t}
   K_{c}(z_1,z_2,z_3)&{}= 
   \dfrac{\beta^2}{2}\Delta(z_1)\Delta(z_2)\Delta(z_3) \widetilde K_{c}(\circ,\circ,\circ)
   +\beta\Delta(z_1)\Delta(z_2+z_3) \widetilde K_{c}(\circ,z_2,z_3)
   \\
   &{}+\beta\Delta(z_2)\Delta(z_3+z_1) \widetilde K_{c}(z_1,\circ,z_3)
   +\beta\Delta(z_3)\Delta(z_1+z_2) \widetilde K_{c}(z_1,z_2,\circ)
   \nonumber\\
   &{}+\Delta(z_1+z_2+z_3) \widetilde K_{c}(z_1,z_2,z_3).
   \nonumber
\end{align}
It includes five contributions which always can be distinguished by the different $\Delta$ factors. The first contribution is expressed directly by the spectral densities of the (\ref{SD_3t_cont0}) type
\begin{equation}\label{Kc0vsSD}
   \widetilde K_{c}(\circ,\circ,\circ)=\tilde I_{ABC}(\circ,\circ,\circ)+\tilde I_{CBA}(\circ,\circ,\circ),
\end{equation}
the next three contributions are expressed in terms of the spectral densities of the (\ref{SD_3t_cont1})--(\ref{SD_3t_cont0}) type
\begin{align}\label{Kc1vsSD}
   \widetilde K_{c}(\circ,z_2,-z_2)&{}=\int\limits_{-\infty}^{+\infty}\rd x_2
       \dfrac{\tilde I_{ABC}(\circ,x_2,-x_2)-\tilde I_{CBA}(-x_2,x_2,\circ)}{x_2-z_2}\\
       &{}-\dfrac1{z_2}\bigl[\tilde I_{ABC}(\circ,\circ,\circ)-\tilde I_{CBA}(\circ,\circ,\circ)\bigr],
       \nonumber
\end{align}
\begin{align}\label{Kc2vsSD}
   \widetilde K_{c}(-z_3,\circ,z_3)&{}=\int\limits_{-\infty}^{+\infty}\rd x_3
       \dfrac{\tilde I_{ABC}(-x_3,\circ,x_3)\re^{\beta x_3}-\tilde I_{CBA}(x_3,\circ,-x_3)\re^{-\beta x_3}}{x_3-z_3}\\
       &{}+\dfrac1{z_3}\bigl[\tilde I_{ABC}(\circ,\circ,\circ)-\tilde I_{CBA}(\circ,\circ,\circ)\bigr],
       \nonumber
\end{align}
and
\begin{align}\label{Kc3vsSD}
   \widetilde K_{c}(z_1,-z_1,\circ)&{}=\int\limits_{-\infty}^{+\infty}\rd x_1
       \dfrac{\tilde I_{ABC}(x_1,-x_1,\circ)-\tilde I_{CBA}(\circ,-x_1,x_1)}{x_1-z_1}\\
       &{}-\dfrac1{z_1}\bigl[\tilde I_{ABC}(\circ,\circ,\circ)-\tilde I_{CBA}(\circ,\circ,\circ)\bigr],
       \nonumber
\end{align}
and the last contribution is expressed in terms of the spectral densities of the (\ref{SD_3t_cont})--(\ref{SD_3t_cont3}) type
\begin{align}\label{KcvsSD3t_reg}
   \widetilde K_{c}(z_1,z_2,z_3)&{}=
     \int\limits_{-\infty}^{+\infty}\rd x_2
       \dfrac{\tilde I_{ABC}(\circ,x_2,-x_2)(1-\re^{-\beta x_2})
             +\tilde I_{CBA}(-x_2,x_2,\circ)(1-\re^{\beta x_2})}{(z_2-x_2)(z_3+x_2)}
     \\
     &{}+\int\limits_{-\infty}^{+\infty}\rd x_3
       \dfrac{\tilde I_{ABC}(-x_3,\circ,x_3)(\re^{\beta x_3}-1)
             +\tilde I_{CBA}(x_3,\circ,-x_3)(\re^{-\beta x_3}-1)}{(z_3-x_3)(z_1+x_3)}
     \nonumber\\
     &{}+\int\limits_{-\infty}^{+\infty}\rd x_1
       \dfrac{\tilde I_{ABC}(x_1,-x_1,\circ)(1-\re^{-\beta x_1})
             +\tilde I_{CBA}(\circ,-x_1,x_1)(1-\re^{\beta x_1})}{(z_1-x_1)(z_2+x_1)}
     \nonumber\\
     &{}-\int\limits_{-\infty}^{+\infty}\rd x_3\int\limits_{-\infty}^{+\infty}\rd x_1
       \dfrac{\tilde I_{ABC}(x_1,-x_3-x_1,x_3)+\tilde I_{CBA}(x_3,-x_3-x_1,x_1)}{(x_3-z_3)(x_1-z_1)}
     \nonumber\\
     &{}-\int\limits_{-\infty}^{+\infty}\rd x_1\int\limits_{-\infty}^{+\infty}\rd x_2
       \dfrac{\tilde I_{ABC}(x_1,x_2,-x_1-x_2)\re^{-\beta x_1}
             +\tilde I_{CBA}(-x_1-x_2,x_2,x_1)\re^{\beta x_1}}{(x_1-z_1)(x_2-z_2)}
     \nonumber\\
     &{}-\int\limits_{-\infty}^{+\infty}\rd x_2\int\limits_{-\infty}^{+\infty}\rd x_3
       \dfrac{\tilde I_{ABC}(-x_2-x_3,x_2,x_3)\re^{\beta x_3}
             +\tilde I_{CBA}(x_3,x_2,-x_1-x_2)\re^{-\beta x_3}}{(x_2-z_2)(x_3-z_3)}.
     \nonumber
\end{align}
Here, we have used the cyclic permutation identities (\ref{CF3tCicle}) according to which there are only two nonidentical three-time correlation functions, e.g. $I_{ABC}(\omega_1,\omega_2,\omega_3)$ and $I_{CBA}(\omega_3,\omega_2,\omega_1)$. Eqs.~(\ref{KcvsSD3t})--(\ref{KcvsSD3t_reg}) give complete representation of the three-time temperature Green functions in terms of the spectral densities.

Now we pass to the solution of the reverse problem~--- finding of the spectral densities from the known multitime temperature Green functions. According to Eqs.~(\ref{KcvsSD3t})--(\ref{KcvsSD3t_reg}) each of five contributions in the three-time Green's function (\ref{KcvsSD3t}) has different set of branch cuts, that allows to extract all ten spectral densities, which enter. 

First of all we perform an analytic continuation of Eq.~(\ref{Kc1vsSD}) ($z_2\to\omega_2\pm\ri\delta$, $z_3=-z_2$, $z_1=0$):
\begin{equation}\label{KcI1}
   \widetilde K_{c}(\circ,z_2,-z_2)\biggr|_2=-\widetilde K_{c}(\circ,-z_3,z_3)\biggr|_3=
     \delta(\omega_2)\mathfrak K_0(\circ,\circ,\circ)+\mathfrak K_{1}(\circ,\omega_2,-\omega_2),
\end{equation}
where
\begin{align}\label{KcI1a}
     \mathfrak K_0(\circ,\circ,\circ)&{}=\tilde I_{ABC}(\circ,\circ,\circ)-\tilde I_{CBA}(\circ,\circ,\circ),\\
     \mathfrak K_{1}(\circ,\omega_2,-\omega_2)&{}=\tilde I_{ABC}(\circ,\omega_2,-\omega_2)
                                    -\tilde I_{CBA}(-\omega_2,\omega_2,\circ).
      \label{KcI1b}
\end{align}
Next, we perform an analytic continuation of Eq.~(\ref{Kc2vsSD}) ($z_3\to\omega_3\pm\ri\delta$, $z_1=-z_3$, $z_2=0$)
\begin{align}\label{KcI2}
  & \widetilde K_{c}(-z_3,\circ,z_3)\biggr|_3=-\widetilde K_{c}(z_1,\circ,-z_1)\biggr|_1=
     -\delta(\omega_3)\mathfrak K_0(\circ,\circ,\circ)+\mathfrak K_{2}(-\omega_3,\circ,\omega_3),\\
  & \mathfrak K_{2}(-\omega_3,\circ,\omega_3)=\tilde I_{ABC}(-\omega_3,\circ,\omega_3)\re^{\beta \omega_3}
           -\tilde I_{CBA}(\omega_3,\circ,-\omega_3)\re^{-\beta \omega_3},
     \label{KcI2b}
\end{align}
and of Eq.~(\ref{Kc3vsSD}) ($z_1\to\omega_1\pm\ri\delta$, $z_2=-z_1$, $z_3=0$)
\begin{align}\label{KcI3}
  & \widetilde K_{c}(z_1,-z_1,\circ)\biggr|_1=-\widetilde K_{c}(-z_2,z_2,\circ)\biggr|_2=
     \delta(\omega_1)\mathfrak K_0(\circ,\circ,\circ)+\mathfrak K_{3}(\omega_1,-\omega_1,\circ),\\
  & \mathfrak K_{3}(\omega_1,-\omega_1,\circ)=\tilde I_{ABC}(\omega_1,-\omega_1,\circ)
              -\tilde I_{CBA}(\circ,-\omega_1,\omega_1).
     \label{KcI3b}
\end{align}
One can see, that in Eqs.~(\ref{KcI1}), (\ref{KcI2}), and (\ref{KcI3}) all factors at $\delta$-functions are the same.

Next step is more complicated and requires a two-stage procedure. First of all we perform an analytic continuation of Eq.~(\ref{KcvsSD3t_reg}) over first frequency $z_1\to\omega_1\pm\ri\delta$ ($z_3=-\omega_1-z_2$)
\begin{align}
   \widetilde K_{c}(z_1,z_2,z_3)\biggr|_1&{}=
       \dfrac{1}{z_2}\bigl[\tilde I_{ABC}(\omega_1,\circ,-\omega_1)(\re^{-\beta \omega_1}-1)
             +\tilde I_{CBA}(-\omega_1,\circ,\omega_1)(\re^{\beta \omega_1}-1)\bigr]
     \\
       &{}+\dfrac{1}{z_2+\omega_1}\bigl[\tilde I_{ABC}(\omega_1,-\omega_1,\circ)(\re^{-\beta \omega_1}-1)
             +\tilde I_{CBA}(\circ,-\omega_1,\omega_1)(\re^{\beta \omega_1}-1)\bigr]
     \nonumber\\
      &{}+\int\limits_{-\infty}^{+\infty}\rd x_2
        \dfrac{\tilde I_{ABC}(\omega_1,x_2,-\omega_1-x_2)(\re^{-\beta \omega_1}-1)
             +\tilde I_{CBA}(-\omega_1-x_2,x_2,\omega_1)(\re^{\beta \omega_1}-1)}{z_2-x_2}
      \nonumber
\end{align}
and then an analytic continuation over second one $z_2\to\omega_2\pm\ri\delta$ ($z_3\to-\omega_1-\omega_2\mp\ri\delta$)
\begin{align}
   \widetilde K_{c}(z_1,z_2,z_3)\biggr|_1\biggr|_2&{}=-\widetilde K_{c}(z_1,z_2,z_3)\biggr|_1\biggr|_3\\
      &{}=\delta(\omega_2)\mathfrak K_{1,2}^{(2)}(-\omega_3,\circ,\omega_3)
      +\delta(\omega_3)\mathfrak K_{1,2}^{(3)}(\omega_1,-\omega_1,\circ)
      +\mathfrak K_{1,2}(\omega_1,\omega_2,\omega_3),
     \nonumber
\end{align}
where
\begin{align}
   \mathfrak K_{1,2}^{(2)}(-\omega_3,\circ,\omega_3)&{}=
     - \tilde I_{ABC}(-\omega_3,\circ,\omega_3)(\re^{\beta \omega_3}-1)
             -\tilde I_{CBA}(\omega_3,\circ,-\omega_3)(\re^{-\beta \omega_3}-1),
     \label{Kc12a}\\
   \mathfrak K_{1,2}^{(3)}(\omega_1,-\omega_1,\circ)&{}=
       -\tilde I_{ABC}(\omega_1,-\omega_1,\circ)(\re^{-\beta \omega_1}-1)
             -\tilde I_{CBA}(\circ,-\omega_1,\omega_1)(\re^{\beta \omega_1}-1),
     \label{Kc12b}\\
   \mathfrak K_{1,2}(\omega_1,\omega_2,\omega_3)&{}=
      -\tilde I_{ABC}(\omega_1,\omega_2,\omega_3)(\re^{-\beta \omega_1}-1)
             -\tilde I_{CBA}(\omega_3,\omega_2,\omega_1)(\re^{\beta \omega_1}-1).
     \label{Kc12c}
\end{align}
We can perform another sequence of the analytic continuations: $z_2\to\omega_2\pm\ri\delta$ ($z_1=-\omega_2-z_3$) and $z_3\to\omega_3\pm\ri\delta$ ($z_1\to-\omega_2-\omega_3\mp\ri\delta$), that gives a different set of relations
\begin{align}
   \widetilde K_{c}(z_1,z_2,z_3)\biggr|_2\biggr|_3&{}=-\widetilde K_{c}(z_1,z_2,z_3)\biggr|_2\biggr|_1\\
      &{}=\delta(\omega_3)\mathfrak K_{2,3}^{(3)}(\omega_1,-\omega_1,\circ)
      +\delta(\omega_1)\mathfrak K_{2,3}^{(1)}(\circ,\omega_2,-\omega_2)
      +\mathfrak K_{2,3}(\omega_1,\omega_2,\omega_3),
      \nonumber
\end{align}
where
\begin{align}
   \mathfrak K_{2,3}^{(3)}(\omega_1,-\omega_1,\circ)&{}=
     -  \tilde I_{ABC}(\omega_1,-\omega_1,\circ)(1-\re^{-\beta \omega_1})
             -\tilde I_{CBA}(\circ,-\omega_1,\omega_1)(1-\re^{\beta \omega_1}),
     \label{Kc23a}\\
   \mathfrak K_{2,3}^{(1)}(\circ,\omega_2,-\omega_2)&{}=
       -\tilde I_{ABC}(\circ,\omega_2,-\omega_2)(\re^{-\beta \omega_2}-1)
             -\tilde I_{CBA}(-\omega_2,\omega_2,\circ)(\re^{\beta \omega_2}-1),
     \label{Kc23b}\\
   \mathfrak K_{2,3}(\omega_1,\omega_2,\omega_3)&{}=
      -\tilde I_{ABC}(\omega_1,\omega_2,\omega_3)(\re^{\beta \omega_3}-\re^{-\beta \omega_1})
             -\tilde I_{CBA}(\omega_3,\omega_2,\omega_1)(\re^{-\beta \omega_3}-\re^{\beta \omega_1}).
     \label{Kc23c}
\end{align}
One can see that
\begin{equation}
   \widetilde K_{c}(z_1,z_2,z_3)\biggr|_1\biggr|_2\neq \widetilde K_{c}(z_1,z_2,z_3)\biggr|_2\biggr|_1,
\end{equation}
but there are common elements in final expressions, e.g.
\begin{equation}
   \mathfrak K_{1,2}^{(3)}(\omega_1,-\omega_1,\circ) + \mathfrak K_{2,3}^{(3)}(\omega_1,-\omega_1,\circ)=0,
\end{equation}
and the following Jacobi type identity is fullfilled
\begin{equation}
    \widetilde K_{c}(z_1,z_2,z_3)\biggr|_1\biggr|_2
   + \widetilde K_{c}(z_1,z_2,z_3)\biggr|_2\biggr|_3
   + \widetilde K_{c}(z_1,z_2,z_3)\biggr|_3\biggr|_1 =0.
\end{equation}
Due to this identity the procedure is unambiguous and resulting set of equations for the spectral densities is not overdetermined.

In general, the three-time temperature Green's functions~(\ref{KcvsSD3t}), as well as multitime Green functions of higher order, have very complicated analytic properties as function of the complex frequencies $z_i$, but they can be separated into contributions with different $\Delta$-factors and different frequency dependences. Analytic properties of this contributions can be also complicated, nevertheless, we can always derive a complete set of equations for the multitime spectral densities from the multitime temperature Green functions by the different sequences of the analytic continuations $z_i\to\omega_i\pm\ri\delta$ accompanied by the constraint~(\ref{z_constr}).

Finally, we find from Eqs.~(\ref{Kc0vsSD}) and (\ref{KcI1a}):
\begin{align}
   \tilde I_{ABC}(\circ,\circ,\circ)=\dfrac12\bigl[\widetilde K_{c}(\circ,\circ,\circ)
                                                      +\mathfrak K_0(\circ,\circ,\circ)\bigr],\\
   \tilde I_{CBA}(\circ,\circ,\circ)=\dfrac12\bigl[\widetilde K_{c}(\circ,\circ,\circ)
                                                      -\mathfrak K_0(\circ,\circ,\circ)\bigr],
   \nonumber
\end{align}
from Eqs.~(\ref{KcI1b}) and (\ref{Kc23b}):
\begin{align}
   & \tilde I_{ABC}(\circ,\omega_2,-\omega_2)=
        \dfrac{\mathfrak K_1(\circ,\omega_2,-\omega_2)}{1-\re^{-\beta\omega_2}}
      + \dfrac{\mathfrak K_{2,3}^{(1)}(\circ,\omega_2,-\omega_2)\re^{-\beta\omega_2}}{(1-\re^{-\beta\omega_2})^2},\\
   & \tilde I_{CBA}(-\omega_2,\omega_2,\circ)=
        \dfrac{\mathfrak K_1(\circ,\omega_2,-\omega_2)}{\re^{\beta\omega_2}-1}
      + \dfrac{\mathfrak K_{2,3}^{(1)}(\circ,\omega_2,-\omega_2)\re^{\beta\omega_2}}{(\re^{\beta\omega_2}-1)^2},
   \nonumber
\end{align}
from Eqs.~(\ref{KcI2b}) and (\ref{Kc12a}):
\begin{align}
   & \tilde I_{ABC}(-\omega_3,\circ,\omega_3)=
        \dfrac{\mathfrak K_2(-\omega_3,\circ,\omega_3)}{\re^{\beta\omega_3}-1}
      + \dfrac{\mathfrak K_{1,2}^{(2)}(-\omega_3,\circ,\omega_3)}{(\re^{\beta\omega_3}-1)^2},\\
   & \tilde I_{CBA}(\omega_3,\circ,-\omega_3)=
        \dfrac{\mathfrak K_2(-\omega_3,\circ,\omega_3)}{1-\re^{-\beta\omega_3}}
      + \dfrac{\mathfrak K_{1,2}^{(2)}(-\omega_3,\circ,\omega_3)}{(1-\re^{-\beta\omega_3})^2},
   \nonumber
\end{align}
from Eqs.~(\ref{KcI3b}) and (\ref{Kc12b}) or (\ref{Kc23a}):
\begin{align}
   & \tilde I_{ABC}(\omega_1,-\omega_1,\circ)=
        \dfrac{\mathfrak K_3(\omega_1,-\omega_1,\circ)}{1-\re^{-\beta\omega_1}}
      - \dfrac{\mathfrak K_{1,2}^{(3)}(\omega_1,-\omega_1,\circ)\re^{-\beta\omega_1}}{(1-\re^{-\beta\omega_1})^2},\\
   & \tilde I_{CBA}(\circ,-\omega_1,\omega_1)=
        \dfrac{\mathfrak K_3(\omega_1,-\omega_1,\circ)}{\re^{\beta\omega_1}-1}
      - \dfrac{\mathfrak K_{1,2}^{(3)}(\omega_1,-\omega_1,\circ)\re^{\beta\omega_1}}{(\re^{\beta\omega_1}-1)^2},
   \nonumber
\end{align}
and from Eqs.~(\ref{Kc12c}) and (\ref{Kc23c}):
\begin{align}
   & \tilde I_{ABC}(\omega_1,\omega_2,\omega_3)=
        \dfrac{\mathfrak K_{2,3}(\omega_1,\omega_2,\omega_3)}{(\re^{\beta\omega_2}-1)(\re^{\beta\omega_3}-1)}
      - \dfrac{\mathfrak K_{1,2}(\omega_1,\omega_2,\omega_3)}{(\re^{\beta\omega_3}-1)(1-\re^{-\beta\omega_1})},
   \\
   & \tilde I_{CBA}(\omega_3,\omega_2,\omega_1)=
        \dfrac{\mathfrak K_{2,3}(\omega_1,\omega_2,\omega_3)}{(1-\re^{-\beta\omega_2})(1-\re^{-\beta\omega_3})}
      - \dfrac{\mathfrak K_{1,2}(\omega_1,\omega_2,\omega_3)}{(1-\re^{-\beta\omega_3})(\re^{\beta\omega_1}-1)},
   \nonumber
\end{align}
that complete our main task and find spectral densities for multitime correlation functions from the known multitime temperature Green functions.

\section{Summary}

In conclusion, we have presented a general approach for derivation of the spectral relations for the multitime correlation functions. An analysis of the frequency dependences of their spectral densities is performed with special attention paid to the consideration of the non-ergodic (conserving) contributions. It is shown that such contributions can be treated in a rigorous way using multitime temperature Green functions: representation of the Green functions in terms of the spectral densities and solution of the reverse problem~--- finding of the spectral densities from the known Green functions are given for the case of the three-time bosonic correlation functions.

Generalization for the case of the higher-order multitime correlation functions and for the functions constructed also from the fermionic operators can be done in the same way, but it was not be considered here. Here we can only note that for the higher-order functions, besides non-ergodic terms, which appear at zero value of one frequency and are connected with the conserving of one operator, the one connected with the conserving of products of operators, which appear at zero value of the sum of corresponding frequencies, always exist and they will be the only ``non-ergodic'' contributions for the pure fermionic correlation functions resulting from the presence of the matrix elements like $\langle f|c^{\dagger}|l\rangle\langle l|c|f\rangle$. Moreover, for some models the fermionic correlation and Green functions will contain only such ``non-ergodic'' contributions, e.g. local four-time two-electron Green's function for the Falicov-Kimball model~\cite{shvaika:349} for which only contributions with zero sum of two frequencies exist.

This publication is based on work supported by Award No. UKP2--2697--LV--06 of the U.S. Civilian Research \& Development Foundation (CRDF). I am grateful to Professor I.V. Stasyuk for useful and stimulating discussions.

\label{last@page}

\end{document}